%% file: root.tex
\documentclass[journal,twoside,web]{ieeecolor}

\usepackage{generic}
\usepackage{cite}
\usepackage{amsmath,amssymb,amsfonts}
\usepackage{algorithmic}
\usepackage{graphicx}
\usepackage{textcomp}
\usepackage{subcaption}
\usepackage{algorithm}

\usepackage{graphicx}
\usepackage{epsfig} 
\usepackage{cite}
\usepackage{lcsys}

\newtheorem{theorem}{Theorem}[section]

\newtheorem{proposition}[theorem]{Proposition}
\newtheorem{corollary}[theorem]{Corollary}
\newtheorem{definition}[theorem]{Definition}

\newtheorem{assumption}[theorem]{Assumption}

\begin{document}

\def\BibTeX{{\rm B\kern-.05em{\sc i\kern-.025em b}\kern-.08em
    T\kern-.1667em\lower.7ex\hbox{E}\kern-.125emX}}
\markboth{\journalname, VOL. XX, NO. XX, XXXX 2017}
{Author \MakeLowercase{\textit{et al.}}: Preparation of Papers for IEEE Control Systems Letters (August 2022)}

\title{Robust $\mathcal H_\infty$ control under stochastic requirements: minimizing conditional value-at-risk instead of worst-case performance}

\author{Ervan Kassarian,
        Francesco Sanfedino,
        Daniel Alazard,
        Andrea Marrazza
\thanks{Ervan Kassarian and Andrea Marrazza are with DYCSYT, 3 Avenue Didier Daurat, 31400 Toulouse, France.}
\thanks{Francesco Sanfedino and Daniel Alazard are with Fédération ENAC ISAE-SUPAERO ONERA, Université de Toulouse, 10 Avenue Marc Pélegrin, Toulouse, 31400, France.}
\thanks{Corresponding author's email: ervan.kassarian@dycsyt.com}
}

\maketitle
\thispagestyle{empty}

\begin{abstract}
Conventional robust $\mathcal H_2/\mathcal H_\infty$ control minimizes the worst-case performance, often leading to a conservative design driven by very rare parametric configurations.
To reduce this conservatism while taking advantage of the stochastic properties of Monte Carlo sampling and its compatibility with parallel computing, we introduce an alternative paradigm that optimizes the controller with respect to a stochastic criterion, namely the conditional value at risk. We present the problem formulation and discuss several open challenges toward a general synthesis framework. The potential of this approach is illustrated on a mechanical system, where it significantly improves overall performance by tolerating some degradation in very rare worst-case scenarios.
\end{abstract}

\begin{IEEEkeywords}
Robust $\mathcal H_\infty$ control, stochastic optimization, risk-aware control, uncertain linear systems. 
\end{IEEEkeywords}

\input{1_introduction}

\input{2_statement}

\input{3_properties}

\input{5_challenges}
\input{4_application}

\section{Conclusion}

To reduce the conservatism of conventional robust $\mathcal H_\infty$ control, which minimizes a worst-case performance that may be very unlikely, we introduced an alternative stochastic approach based on the conditional value at risk.
Its capacity to significantly improve the performance of uncertain control systems, at the expense of tolerating rare degrading worst-case configurations, was illustrated on a mechanical system. Finally, we presented a first step toward a synthesis method and highlighted several remaining open challenges.
 
\bibliographystyle{IEEEtran}
\bibliography{library}

\end{document}

%% file: 1_introduction.tex
\section{Introduction}
\IEEEPARstart{R}{obust} structured $\mathcal H_\infty$ control \cite{apkarian2017} traditionally aims at finding a controller that minimizes the worst-case performance across all possible parametric configurations. Denoting as $T_{zw}(\mathrm s, k,\delta)$ a transfer function from input $w$ to output $z$, which depends on the vectors $k$ of decision variables and $\delta$ of uncertain parameters, this formulation reads
\begin{equation} \label{eq:Hinfproblem}
 \underset{k \in \mathcal K}{\text{minimize }} \underset{\delta \in \mathcal D_\delta}{\max } \; || T_{zw}(\mathrm s, k,\delta) ||_\infty \;.
\end{equation}
Here, $\mathcal K$ denotes the admissible space, describing the controllers with a chosen structure (fixed-order, PID, etc.), and $\mathcal D_\delta$ is generally the hypercube $[-1,1]^m$ after normalization of the $m$ uncertain parameters. The case of a nonlinear constraint restricting $\mathcal D_\delta$ was also addressed in~\cite{Kassarian2025}.
When there are multiple control objectives, it is sought to solve:
\begin{equation} \label{eq:Hinf_multiobj}
 \begin{array}{lll}
    & \underset{k \in \mathcal K}{\text{minimize}} & \underset{\delta \in \mathcal D_\delta}{\text{max}}  \; || T_{z_1w_1}(s, k,\delta) ||  \\
    & \text{subject to } &  \underset{2 \leq i \leq n }{\text{max}} \; \underset{\delta \in \mathcal D_\delta}{\text{max}}  \; || T_{z_iw_i}(s, k,\delta) || \leq 1 
 \end{array}
\end{equation}
where $|| \cdot ||$ may refer to the $\mathcal H_2$ or $\mathcal H_\infty$ system norm. The channel $w_1 \rightarrow z_1$ corresponds to a \textit{soft} requirement to be minimized, while the other channels $w_i \rightarrow z_i$ ($i>1$) are \textit{hard} requirements, that must be less than 1 (once normalized) but not necessarily minimized. 
As this formulation does not involve any notion of probability, the worst-case configuration is often extremely unlikely, leading to a conservative design with regard to industrial requirements that are generally defined in a stochastic sense and relying on a probabilistic description of the uncertain parameters (e.g. gaussian, uniform, or other distributions).

When considering deterministic parameters $\delta \in [-1,1]^m$, the state of the art for robust $\mathcal H_\infty$ control is the nonsmooth optimization algorithm of~\cite{Apkarian2015a}, implemented in MATLAB's function \textsc{systune} \cite{systune}, that solves the inner maximization problem of \eqref{eq:Hinfproblem} using a local search enabled by the computation of Clarke subdifferentials~\cite{Clarke90} of the $\mathcal H_\infty$ norm~\cite{Boyd91}.
Identified worst-case configurations are iteratively added to an active set $\mathcal D_a \subset \mathcal D_\delta$ on which the outer minimization problem is solved based on the algorithm of~\cite{Apkarian2006}.
Then, global robustness verification can be done for example with Monte Carlo sampling~\cite{Tempo1996} or with $\mu$-analysis~\cite{doyle1982}. The former provides statistical guarantees based on the worst observed sample, while the latter, generally based on branch-and-bound algorithms~\cite{roos2013systems}, provides a formal guarantee of detecting the global worst-case configuration. However, as noted above, this worst-case scenario can be extremely rare, which motivated the development of probabilistic $\mu$-analysis~\cite{Biannic2021}.
Nevertheless, $\mu$-based techniques exhibit limited scalability with the number of parameters, since they rely on bounding the structured singular value, whose exact computation is NP-hard \cite{Toker98}. Furthermore, they are not oriented toward robust controller synthesis, which is the primary concern of the present work.

This paper proposes an alternative paradigm for robust $\mathcal H_2/\mathcal H_\infty$ controller synthesis with two main objectives. The first is to optimize a robust controller in the sense of stochastic requirements, better aligning the synthesis framework with the underlying uncertainty-generating process and thereby reducing conservatism. The second is to leverage Monte Carlo methods, which are commonly used in stochastic programming~\cite{Shapiro2009}, provide stochastic guarantees, and naturally exploit parallel computing.

For this, we rely on the conditional value at risk (CVAR),
a common concept in the field of finance to quantify potential extreme losses in a portfolio. The rationale for employing CVAR over other risk measures, such as the quantile, is twofold. First, it captures both the quantile threshold and the severity of losses beyond it, providing a more complete measure of risk.
Second, it is particularly suitable for optimization, as established in the literature \cite{Rockafellar2000}.
This formulation is introduced in Section~\ref{sec:2}, and adapted in Section~\ref{sec:3} to be more suitable for optimization. Section~\ref{sec:optim} presents a first step toward a solution method and identifies several open challenges that remain to be addressed. Finally, Section~\ref{sec:4} illustrates this method on the robust control of a mechanical system.


%% file: 2_statement.tex
\section{Problem statement}

\label{sec:2}

\begin{figure}
\centering
\includegraphics[width=.4\linewidth]{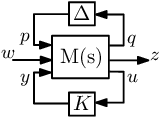}
\caption{LFR representation of $T_{zw}(\mathrm s, k,\delta)$}
\label{fig:lft}
\end{figure}

Considering a vector $\delta \in \mathbb R^m$ of uncertain parameters and a controller $K$ structured by a vector $k \in \mathcal K$ of decision variables (e.g. fixed-order controller, observer-based, PID, etc.), let us consider a transfer function $T_{zw}(\mathrm s, k,\delta)$ under the Linear Fractional Representation (LFR) as in Fig.~\ref{fig:lft},
where $\Delta = \textrm{diag} (\delta_i I_{n_i})$ ($n_i$ is the number of repetitions of $\delta_i$). 

\vspace{1mm}

\begin{assumption} \label{assumption_bounded}
    The random variable $\delta$ has bounded support $\mathcal D_\delta$, and its probability distribution admits a density $p(\delta)$.
\end{assumption}

\vspace{1mm}

Assuming a bounded support is common in control engineering, as it is necessary in the traditional deterministic framework to define a worst case (as in Eq.~\eqref{eq:Hinfproblem}). In this paper, it is necessary to realistically make Assumption~\ref{assumption_stability} below. Without loss of generality, we consider that $\mathcal D_\delta \subset [-1,1]^m$ and that $\delta=0$ corresponds to the nominal configuration.

For a system norm $|| \cdot ||$ which may be the $\mathcal H_2$ or $\mathcal H_\infty$ norm\footnote{
For the $\mathcal H_2$ norm, we assume that the matrix $D$ of the state-space representation is such the $\mathcal H_2$ norm is correctly defined, cf. e.g. \cite{Rautert1997,apkarian2009,Arzelier2010}.
} in this paper, the \textit{loss function}:
\begin{equation} \label{eq:loss}
    L_{zw}(k, \delta) = || T_{zw}(\mathrm s, k,\delta) ||
\end{equation}
is a random variable that depends on the decision vector $k$.

\vspace{1mm}

\begin{assumption} \label{assumption_stability}
    Given $k \in \mathcal K$, the loss function $L_{zw}(k, \delta)$ is said to satisfy Assumption~\ref{assumption_stability} when $T_{zw}(\mathrm s, k,\delta)$ is internally stable for all $\delta \in \mathcal D_\delta$.
\end{assumption}

\vspace{1mm}

Assumption~\ref{assumption_stability} ensures that the loss function is well-defined for any $\delta$, and thus allows us to define the expected value in the following. This assumption might be seen as restrictive; but in practice, it can anyway be desirable to guarantee stability even in rare worst-case scenarios, with the following stochastic formulation relaxing only the performance requirements measured with the $\mathcal H_\infty$ or $\mathcal H_2$ norms.

\vspace{1mm}

\begin{assumption} \label{assumption}
    Given $k \in \mathcal K$, the loss function  $L_{zw}(k, \delta)$ is said to satisfy Assumption~\ref{assumption} when its distribution function $\Psi^{zw}(k, \alpha)$ is continuous.
\end{assumption}

\vspace{1mm}

Assumption~\ref{assumption} is adopted here following the work of \cite{Rockafellar2000} to properly introduce Definition~\ref{def:var}.


\vspace{4mm}

\begin{definition}(\textit{$\beta$-VAR, $\beta$-CVAR}) \label{def:var}

\noindent Given a loss function $L_{zw}(k, \delta)$ satisfying Assumptions~\ref{assumption_bounded}, \ref{assumption_stability} and \ref{assumption}, and $\beta \in [0, 1)$, the \textit{$\beta$-Value at risk ($\beta$-VAR)}, noted $\text{VAR}_\beta^{zw}(k)$, is:
\begin{equation}
    \text{VAR}_\beta^{zw}(k) = \text{min} \lbrace \alpha \in \mathbb R: \Psi^{zw}(k, \alpha) \geq \beta \rbrace \;.
\end{equation}
The $\beta$-Conditional value at risk ($\beta$-CVAR) is:
\begin{equation} \label{eq:betaCVAR}
    \textrm{CVAR}_\beta^{zw}(k) = \frac{1}{1-\beta}\int_{L_{zw}(k,\delta) \geq \textrm{VAR}_\beta^{zw}(k) } L_{zw}(k, \delta) p (\delta) d\delta \; .
\end{equation}
\end{definition}

\vspace{1mm}

\begin{figure}
\includegraphics[width=\linewidth]{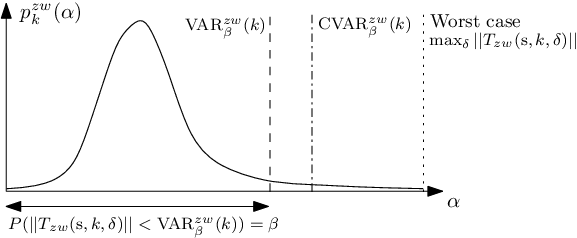}
\caption{Probability density $p^{zw}_k(\alpha)$ of the loss function $|| T_{zw}(\mathrm s, k,\delta) ||$ with $\beta$-VAR, $\beta$-CVAR and worst case}
\label{fig:schema_distribution}
\end{figure}

In other words, when looking at the probability distribution of the loss function $L_{zw}(k, \delta)$, $\text{VAR}_\beta^{zw}(k)$ separates the $\beta$ most favorable cases from the $(1-\beta)$ worst cases, while $\textrm{CVAR}_\beta^{zw}(k)$ represents the conditional expected value of the loss associated with the worst cases $\delta$ relative to that loss being greater than~$\text{VAR}_\beta^{zw}(k)$. This is schematically represented in Fig.~\ref{fig:schema_distribution}. 

In this paper, we focus on the problem:
\begin{equation} \label{eq:HinfCVAR_multiobj}
 \begin{array}{lll}
    &  \underset{k \in \mathcal K}{\text{minimize }} & \textrm{CVAR}_{\beta_1}^{z_1w_1}(k) \\
    & \text{subject to } &  \textrm{CVAR}^{z_iw_i}_{\beta_i}(k) \leq 1 \;, \; 2 \leq i \leq n 
 \end{array}
\end{equation}
which replaces problem~\eqref{eq:Hinf_multiobj}, with the main differences: (i) the parameters $\delta$ are defined as random variables rather than deterministic parameters, over the same bounded support (Assumption~\ref{assumption_bounded}), (ii) instead of minimizing the worst case without considering any probability, we seek to minimize the conditional expected value of the ($1-\beta_i$) worst cases.


\section{Computation of VAR and CVAR}
\label{sec:3}

\subsection{Formulation as an optimization problem}

Definition~\ref{def:var} does not allow for a straightforward computation of the VAR and CVAR. Instead, one typically relies on the following result of Rockafellar~\cite{Rockafellar2000}, in which both are characterized as solutions of a convex optimization problem.

\vspace{1mm}

\begin{proposition} (\textit{Expression of $\beta$-VAR and $\beta$-CVAR as solutions of a convex minimization problem} \protect{\cite{Rockafellar2000})} \label{prop:betavar}

\noindent Under Assumptions~\ref{assumption_bounded}, \ref{assumption_stability} and \ref{assumption}, let us define the function
\begin{equation} \label{eq:F_beta}
    F^{zw}_\beta(k, \alpha) = \alpha + \frac{1}{1-\beta} \int_{\delta \in \mathbb R^m} [L_{zw}(k, \delta) - \alpha]_+ p (\delta) d\delta
\end{equation}
where $[x]_+ = \text{max}(x, 0)$.
The $\beta$-CVAR of the loss function $L_{zw}(k,\delta)$ is the minimum of $F_\beta(k, \alpha)$ as a function of $\alpha$:
\begin{equation} \label{eq:CVAR_min}
    \textrm{CVAR}_\beta^{zw}(k) = \min_{\alpha \in \mathbb R} F^{zw}_\beta(k, \alpha)
\end{equation}
and the $\beta$-VAR is the left-end value of the set of all $\alpha$ that attain this minimum (possibly reducing to a single point):
\begin{equation}
    \text{VAR}_\beta^{zw}(k) = \text{left end point of } \arg \min_{\alpha \in \mathbb R} F^{zw}_\beta(k, \alpha) \;.
\end{equation}
Moreover, $F^{zw}_\beta(k, \alpha)$ is convex and continuously differentiable as a function of $\alpha$.
\end{proposition}

\vspace{1mm}

\begin{corollary} \textit{Optimal controller for the $\beta$-CVAR problem}

\noindent Problem~\eqref{eq:HinfCVAR_multiobj} is equivalent to:
\begin{equation} \label{eq:HinfCVAR_multiobj_bis}
 \begin{array}{lll}
    &  \underset{k \in \mathcal K, \alpha \in \mathbb R^n}{\text{minimize }} & F^{z_1w_1}_{\beta_1}(k, \alpha_1) \\
    & \text{subject to } &  F^{z_iw_i}_{\beta_i}(k, \alpha_i) \leq 1 \;, \; 2 \leq i \leq n 
 \end{array}
\end{equation}
where we have gathered all $\alpha_i$ in a vector $\alpha \in \mathbb R^n$.

\end{corollary}


%% file: 3_properties.tex
\subsection{Sample average approximation (SAA)}

The function $F^{zw}_\beta(k, \alpha)$ can be approximated by sampling $N$ i.i.d. realizations $\tilde\delta^1$, $\tilde\delta^2$, ... $\tilde\delta^N$ from the density $p$ and defining the sample average approximation (SAA)~\cite{Homem2014}
\begin{equation} \label{eq:SAA}
    \widetilde F^{zw}_\beta(k, \alpha, \tilde \delta) = \alpha + \frac{1}{(1-\beta)N} \sum_{i=1}^N [L_{zw}(k, \tilde\delta^i) - \alpha]_+
\end{equation}
that converges point-wise to $F^{zw}_\beta(k, \alpha)$ when $N \to \infty$~\cite{Shapiro2003}.

Optimization of $\beta$-CVAR based on the SAA has been addressed in finance-related literature \cite{Rockafellar2000,bardou2009,Tamar2014,Takano2015}.
As an approximation of the stochastic programming problem~\eqref{eq:HinfCVAR_multiobj_bis}, we may solve the deterministic SAA problem:

\begin{equation} \label{eq:HinfCVAR_multiobj_bis_approx}
 \begin{array}{lll}
    &  \underset{k \in \mathcal K, \alpha \in \mathbb R^n}{\text{minimize }} & \widetilde F^{z_1w_1}_{\beta_1}(k, \alpha_1, \tilde \delta) \\
    & \text{subject to } &  \widetilde  F^{z_iw_i}_{\beta_i}(k, \alpha_i, \tilde\delta) \leq 1 \;, \; 2 \leq i \leq n  \\
    && \underset{\delta \in \mathcal D_\delta}{\max } \; \lambda^{\mathcal R}_{\max}(A(k, \delta)) < 0
 \end{array}
\end{equation}
where $A(k, \delta)$ is the dynamics matrix of the state-space representation of the closed loop, and $\lambda^{\mathcal R}_{\max}(\cdot)$ denotes the spectral abscissa. The constraint on the spectral abscissa ensures that Assumption~\ref{assumption_stability} remains satisfied over the whole support $\mathcal D_\delta$ and not only over the samples $\tilde \delta^{(i)}$. How this constraint can be enforced is discussed in Section~\ref{sec:optim}.

\subsection{Confidence interval}

\label{sec:confidence_interval}

To obtain an accurate solution to the original problem, it is necessary to choose $N$ sufficiently large. \cite[Theorem 3]{Cherukuri2024} characterizes the convergence rate and derives the number of samples required to guarantee a prescribed accuracy level with a given confidence. However, this result also depends on the regularity of the uncertain loss function, which is not known a priori. In practice, thanks to the expression of VAR and CVAR as solutions of an optimization problem (Proposition~\ref{prop:betavar}), the procedure proposed in~\cite[Section 3.2]{Mak1999} allows to assess the gap between the estimated $\beta$-CVAR:
\[
\widehat{\textrm{CVAR}}_\beta^{zw}(k) = \min_{\alpha \in \mathbb R} \widetilde F^{zw}_\beta(k, \alpha, \tilde \delta)
\]
and the true one, solution of Eq.~\eqref{eq:CVAR_min}. Specifically, given a candidate solution $\hat \alpha$, one draws $M$ i.i.d. batches of $n$ vectors $\tilde\delta^{i1}$, $\tilde\delta^{i2}$, ... $\tilde\delta^{in}$, $i=1, 2, ...,  M$, and define for each batch:
\[
G_i(n) = \frac{1}{n} \sum_{j=1}^n \widetilde F^{zw}_\beta(k, \hat \alpha, \tilde\delta^{ij}) - \min_{\alpha \in \mathbb R} \frac{1}{n} \sum_{j=1}^n  \widetilde F^{zw}_\beta(k, \alpha, \tilde\delta^{ij}) \;.
\]
Since the expected value of the left and right terms are respectively an upper and lower bound of the true optimum, each $G_i(n)$ is an observation of a random variable $G(n)$ representing the optimality gap.
Noting $\bar G(M,n) = \frac{1}{M} \sum_{i=1}^M G_i(n)$, \cite{Mak1999} shows that 
$ \sqrt M (\bar G(M,n) - E(G(n))) $ converges in distribution to a normal law.
Noting $\sigma^2_G(M,n)$ the sample variance estimator of the variance of $G(n)$, and $t_{M-1,\gamma}$ the $(1-\gamma)$-quantile of a Student $t$ distribution with $M-1$ degrees of freedom, then
\begin{equation} \label{eq:confidence}
\left[ 0, \; \bar G (M,n) + \frac{t_{M-1,\gamma} \sigma_G(M,n)}{\sqrt M} \right]
\end{equation}
is a $(1-\gamma)$-interval of confidence for the optimality gap \cite{Mak1999}.

%% file: 5_challenges.tex
\section{Optimization}

\subsection{Optimization procedure}

\label{sec:optim}

Based on the formulation of previous sections, Algorithm~\ref{algo} is proposed to tackle the stochastic optimization problem~\eqref{eq:HinfCVAR_multiobj}.

\begin{algorithm}
\caption{CVAR optimization}\label{algo}
\textbf{Initialization}: Active configurations $\mathcal D_\delta^a = \lbrace 0 \rbrace$, number of samples $N$, tolerance $\epsilon$.

\vspace{1mm}
\textbf{Step 1:} Find $\hat k_0$ that robustly stabilizes $T_{zw}(s, k, \delta)$ over the whole support $\mathcal D_\delta$, and update $\mathcal D_\delta^a$.

\vspace{1mm}
\textbf{Step 2:} Form the SAAs with $N$ samples, initialize $k=\hat k_0$ and solve the SAA problem~\eqref{eq:HinfCVAR_multiobj_bis_approx} where robust stability over the whole support:
\[
\underset{\delta \in \mathcal D_\delta}{\max} \; \lambda^{\mathcal R}_{\max}(A(k, \delta)) < 0
\]
is replaced with stability over the active configurations:
\[
\underset{\delta \in \mathcal D_\delta^a}{\max} \; \lambda^{\mathcal R}_{\max}(A(k, \delta)) < 0 \;.
\]
The optimal controller is noted $\hat k$.

\vspace{1mm}
\textbf{Step 3:} Perform worst-case search in stability:
\[
\delta^* = \textrm{arg } \underset{\delta \in \mathcal D_\delta}{\max} \; \lambda^{\mathcal R}_{\max}(A(\hat k, \delta)) \;.
\]
If $\lambda^{\mathcal R}_{\max}(A(\hat k, \delta^*)) > 0$, add $\delta^*$ to $\mathcal D_\delta^a$ and go back to step~2. Otherwise, go to step 4.

\vspace{1mm}
\textbf{Step 4:} Evaluate the CVARs of problem~\eqref{eq:HinfCVAR_multiobj_bis_approx}, with sufficiently tight intervals of confidence. If they are different from those obtained in step 2 by more than $\epsilon$\%, increase $N$ and go back to step~2. Otherwise, terminate.
\end{algorithm}

Step~1 allows to provide an initial controller with robust stability over the whole support, as required by Assumption~\ref{assumption_stability}. It can be performed with~\cite{Apkarian2015a}, which relies on the algorithm of~\cite{Apkarian2006} and the active configurations approach described in next paragraph. In~\cite{Apkarian2006}, the pure stabilization problem is addressed with the shifted $\mathcal H_\infty$ norm~\cite{Boyd91}. 
Optimization of the SAA approximation is performed in step~2; it can be addressed with deterministic nonsmooth optimization, cf. section~\ref{sec:challenges}.
A worst-case search for unstable configurations is performed in step~3, as discussed in next paragraph. Finally, step~4 evaluates the estimated CVARs obtained with the optimized controller and their confidence intervals of Section~\ref{sec:confidence_interval} to validate or invalidate the controller. 

The identification of the global worst-case stability configuration being NP-hard \cite{Toker98}, the stability over the whole support $\mathcal D_\delta$ is addressed through the active configurations approach of \cite{Apkarian2015a}. It consists of enforcing stability of configurations belonging to a subset $\mathcal D_\delta^a \subset \mathcal D_\delta$, initialized with the nominal configuration, and updating this subset through (deterministic) worst-case searches (step~3).
This procedure is necessary even when the initial controller stabilizes the system over the whole support (step~1), because this property may not be maintained when solving the SAA problem (step~2).
If a destabilizing configuration $\delta^*$ is identified in step~3, it is added to $\mathcal D_\delta^a$, and step~2 is repeated. Note that the step~1 typically also uses this approach~\cite{Apkarian2015a}, and thus, it also updates $\mathcal D_\delta^a$.

\subsection{Open challenges}

\label{sec:challenges}

While we have proposed a general method to address the CVAR optimization, there are still open challenges to address the formulated problem in a systematic and efficient way.

Future research should focus on developing suitable nonsmooth algorithms to address step~2 of Algorithm~\ref{algo}, efficiently and with convergence certificates.
Of particular interest is the work of \cite{Noll2013}, which proposes an algorithm that tolerates inexact computation of the gradient or of the function itself. This algorithm is specifically designed for lower-$C^1$ functions, a class to which the SAA $\widetilde F^{zw}_\beta(k, \alpha)$ can easily be shown to belong under Assumptions~\ref{assumption_bounded}, \ref{assumption_stability} and \ref{assumption}, since this property holds for $|| T_{zw}(s,k, \delta)||_\infty$ as a function of $k$ as long as $T_{zw}(s,k, \delta)$ is internally stable \cite[Proposition 2]{Apkarian2015a}.
In particular, \cite{Noll2013} illustrates their method on the $\mathcal H_\infty$ norm where the exactness of the computation, based on a dichotomy involving the bounded real lemma, can be relaxed. This property is particularly desirable in our SAA where we have to evaluate many times the $\mathcal H_\infty$ norm, and where our SAA is itself an approximation.

Importance sampling \cite{Homem2014,Aolaritei2025} is a technique used in rare event simulation that can reduce the variance of the SAA and reduce the required number of samples. It may particularly be useful when choosing a $\beta$ very close to 1. For example, it was applied to CVAR in \cite{bardou2009,Sun2009,Tamar2014}.
Performing importance sampling in a context of optimization is challenging, since an appropriate sampling distribution may depend on $k$ and therefore has to be updated during optimization \cite{bardou2009,Aolaritei2025}.

Finally, in this paper, the distributions are assumed to be perfectly known. When this assumption does not hold, distributionally robust control provides a natural extension; see, for example, \cite{VanParys2015,Hakobyan2022}.

%% file: 4_application.tex
\section{Application}

\label{sec:4}

\subsection{Benchmark}

\begin{figure}[h!]
\centering
\includegraphics[width=.7\linewidth]{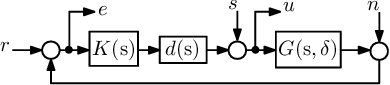}
\caption{Benchmark: closed-loop system}
\label{fig:closedloop}
\end{figure}

We consider the benchmark of \cite{Kassarian2025}, namely the attitude control of a flexible spacecraft $G(s, \delta)$ to follow a reference~$r$.
The closed loop is represented in Fig.~\ref{fig:closedloop}, where $d(s)$ represents actuator dynamics.
The control input $u$ of the plant is the torque (3 components) applied to the satellite and the output vector contains the 3 angle measurements polluted by some noise $n$.
We note $k \in \mathcal K$ the decision variables describing the controllers $K(s)$ of the desired structure: one 4-th order controller per spacecraft axis, each one taking the measured angle and rate as inputs and generating a control torque.
The closed-loop system has 50 states, 43 uncertain parameters $\delta_i$, and 72 decision parameters~$k_i$.

The uncertain parameters represent mechanical properties. Once normalized, they all follow a Gaussian distribution of mean 0 and standard deviation $1/3$, truncated at 3 $\sigma$, except for one parameter (representing an angle) which follows a uniform distribution between -1 and 1.
Additionally, the vectors $\delta$ that do not satisfy a nonlinear constraint $c(\delta) \leq 0$, related to some mechanical constraint detailed in \cite{Kassarian2025}, must be removed from the robustness analysis. In the deterministic setting, this constraint can be addressed in the framework of constrained optimization as in~\cite{Kassarian2025}. In the $\beta$-CVAR formulation, when performing Monte Carlo sampling, samples that do not satisfy the constraint are simply discarded; all following results ($\beta$-VAR, $\beta$-CVAR, mean, distribution) refer to the resulting conditional distribution.

The control problem consists in robustly placing a control bandwidth of 0.1 rad/s (Req.~3) while ensuring a modulus margin of 0.5 (Req.~2) and minimizing the variance of the actuator command in response to measurement noise (Req.~1). It involves the 3 loss functions:
\begin{equation*}
    \begin{aligned}
        L_{un}(s,k,\delta) &= || W_1 T_{n \rightarrow u} (s, K, \delta) ||_2^2\\
        L_{us}(s,k,\delta) &= || W_2 T_{s \rightarrow u} (s, K, \delta) ||_\infty\\
        L_{er}(s,k,\delta) &= || W_3(s) T_{r \rightarrow e} (s, K, \delta) ||_\infty
    \end{aligned}
\end{equation*}
with $W_1$ a normalization constant, $W_2=\frac{1}{2}$ to ensure the desired modulus margin, $W_3(s) = \frac{10 \mathrm s + 1}{2(10 \mathrm s + 0.01)} I_3$ a weighting filter to impose the desired bandwidth. Next sections detail the corresponding deterministic and stochastic formulations.

\subsection{Formulation as a deterministic control problem}

\label{sec:4_deter}

\begin{figure}[ht] 
  \begin{subfigure}[b]{\linewidth}
    \centering
    \includegraphics[width=\linewidth]{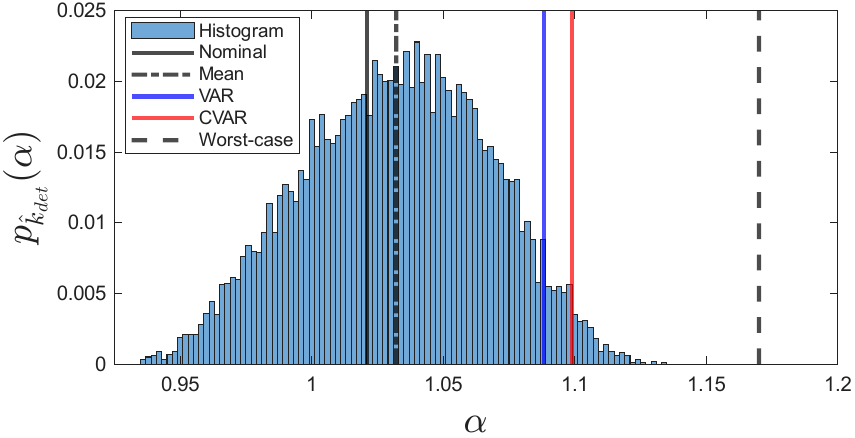} 
    \caption{Density of $L_{un}(s,\hat k_{det},\delta)$ (Req.~1)} 
    \label{fig:distrib_Kinit_req1}
  \end{subfigure} \hfill
  \begin{subfigure}[b]{\linewidth}
    \centering
    \includegraphics[width=\linewidth]{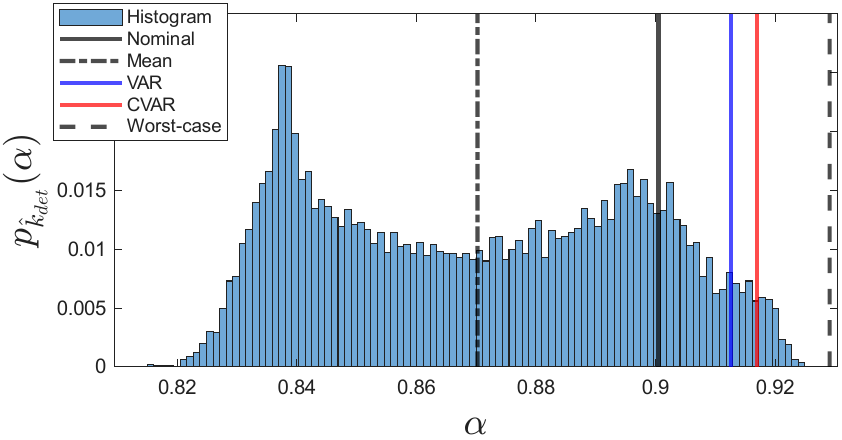} 
    \caption{Density of $L_{us}(s,\hat k_{det},\delta)$ (Req.~2)} 
    \label{fig:distrib_Kinit_req2}
  \end{subfigure} 
  \caption{Empirical probability densities with solution $\hat k_{det}$}
  \label{fig:distrib_Kinit} 
\end{figure}

In the deterministic formulation, the space of (deterministic) uncertain parameters is the support of $\delta$:
\[
   \mathcal D_\delta = \lbrace \delta \in [-1,1]^m \text{ such that } c(\delta) \leq 0 \rbrace 
\]
without any consideration of probability.
Under the traditional robust control framework, the control problem reads:
\begin{equation*} \label{eq:traditional_problem}
\begin{array}{lllr}
    & \underset{K \in \mathcal K}{\text{minimize}} & \underset{\delta \in \mathcal D_\delta}{\max } \; L_{un}(s,k,\delta) & \text{(Req.~1)}\\
    & \text{subject to } &  \underset{\delta \in \mathcal D_\delta}{\max } \; L_{us}(s,k,\delta) \leq 1 & \text{(Req.~2)}\\
    & & \underset{\delta \in \mathcal D_\delta}{\max } \; L_{er}(s,k,\delta) \leq 1  & \text{(Req.~3)}
\end{array}
\end{equation*}
and is solved with the method of \cite{Kassarian2025}. Let us note $\hat k_{det}$ the decision parameters describing the optimal controller.
The resulting densities for $L_{un}(s,\hat k_{det},\delta)$ (Req.~1, Fig~\ref{fig:distrib_Kinit_req1}) and $L_{us}(s,\hat k_{det},\delta)$ (Req.~2, Fig~\ref{fig:distrib_Kinit_req2}) are presented in Fig.~\ref{fig:distrib_Kinit} using 10000 samples. It can be verified that the worst case for Req.~2, computed with the deterministic method of \cite{Kassarian2025}, is indeed less than 1, as imposed. The mean, $\beta$-VaR and $\beta$-CVaR ($\beta$ = 95\%) are reported even though they were not used when solving the optimization problem. Nominal performance, i.e. when $\delta=0$, is also indicated for completeness. All performance metrics are summarized in Table~\ref{table:results}, including Req.~3, which is not represented in the figures.

\subsection{Formulation as a stochastic control problem}

Based on the notations of Definition~\ref{prop:betavar}, we fix $\beta = 95$\% and seek to solve the stochastic problem:
\begin{equation*} \label{eq:stochastic_problem}
\begin{array}{lllr}
    &  \underset{K \in \mathcal K}{\text{minimize }} & \textrm{CVAR}^{un}_\beta(k) & \text{(Req.~1)}\\
    & \text{subject to } &  \textrm{CVAR}^{us}_\beta(k) \leq 1 & \text{(Req.~2)} \\
    &  &  \textrm{CVAR}^{er}_\beta(k) \leq 1 & \text{(Req.~3)} \\
\end{array}
\end{equation*}
which we replace, for a set $\tilde \delta$ of i.i.d. realizations, by the deterministic problem:
\begin{equation*} \label{eq:betaCVAR_problem}
\begin{array}{lllr}
    &  \underset{K \in \mathcal K, \; \alpha \in \mathbb R^3}{\text{minimize }} & \widetilde F^{un}_\beta(k, \alpha_1, \tilde \delta) & \text{(Req.~1)}\\
    & \text{subject to } &  \widetilde F^{us}_\beta(k, \alpha_2, \tilde \delta) \leq 1 & \text{(Req.~2)} \\
    &  &  \widetilde F^{er}_\beta(k, \alpha_3, \tilde \delta) \leq 1 & \text{(Req.~3)}\\
    && \underset{\delta \in \mathcal D_\delta}{\max } \; \lambda^{\mathcal R}_{\max}(A(k, \delta)) < 0
\end{array}
\end{equation*}

We solved this problem with Algorithm~\ref{algo}, using $\hat k_{\textrm{det}}$ as initial controller (step 1).
Step 2 was performed with MATLAB's function \textsc{fmincon} \cite{matlaboptim} using the sequential quadratic programming (SQP) algorithm, even though convergence to an optimum is not formally guaranteed due to the nonsmoothness of the SAA.
The constraint on the spectral abscissa was not explicitly taken into account when solving Step~2, but stability over the whole domain was verified a posteriori in Step~3 using the method of~\cite{Kassarian2025} (note that Req.~2, by imposing a constraint on the CVAR of the stability margins, helps enforce stability).
2500 samples were used initially in the SAAs, and this number was increased to 10000 following step 4. With these 10000 samples, the 99\%  confidence interval of Eq.~\eqref{eq:confidence} was at most $0.0008$ among the 3 requirements, which was considered sufficient for validation.
We note $\hat k_{sto}$ the decision parameters describing the optimal controller. 

\begin{figure}[ht] 
  \begin{subfigure}[b]{\linewidth}
    \centering
    \includegraphics[width=\linewidth]{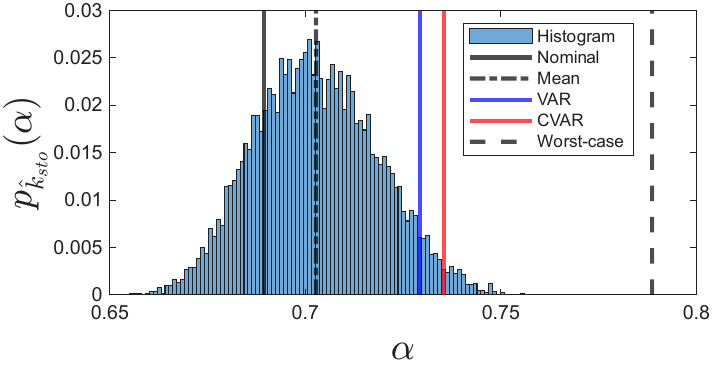} 
    \caption{Density of $L_{un}(s,\hat k_{sto},\delta)$ (Req.~1)} 
    \label{fig:distrib_Kopt_req1}
  \end{subfigure} \hfill
  \begin{subfigure}[b]{\linewidth}
    \centering
    \includegraphics[width=\linewidth]{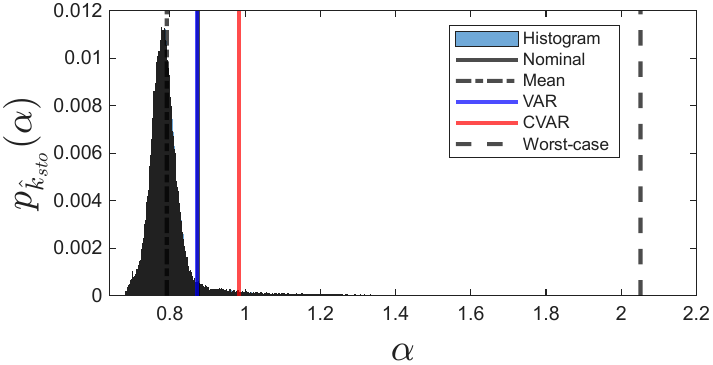} 
    \caption{Density of $L_{us}(s,\hat k_{sto},\delta)$ (Req.~2)} 
    \label{fig:distrib_Kopt_req2}
  \end{subfigure} 
  \caption{Empirical probability densities with solution $\hat k_{sto}$}
  \label{fig:distrib_Kopt} 
\end{figure}

To compare with the results of the deterministic formulation of Section~\ref{sec:4_deter}, we also computed the worst case with the method of~\cite{Kassarian2025}, although this computation was not performed during the optimization itself.
The results are presented in Fig.~\ref{fig:distrib_Kopt} and summarized in Table~\ref{table:results}.
In particular:

\vspace{1mm}

\noindent 1) Fig.~\ref{fig:distrib_Kopt_req2}: The $\beta$-CVAR of $L_{us}(s,\hat k_{sto},\delta)$ (Req.~2) is 0.98, which is less than 1 as imposed. A worst-case of 2.05 (again found with the deterministic method of~\cite{Kassarian2025}) is tolerated, whereas it would have been rejected with the traditional formulation. Nevertheless, this worst case is very rare.
To illustrate this statement, we also mention that among the 10,000 samples used to construct the histogram, the maximum observed value was 1.58. This implies, with 99.99\% confidence, that at least 99.9\% of the underlying distribution lies below 1.58.\footnote{{ For a random function $f(\delta)$, noting $\delta^*$ the sample realizing the largest function evaluation, \cite{Tempo1996} shows that
\begin{equation*} \label{eq:MonteCarlo}
    \mathrm P ( \mathrm P ( f(\delta) > f(\delta^*)) \leq \epsilon ) \geq 1-\gamma
\end{equation*}
is guaranteed when $N$ verifies:
    \begin{equation*}
    N > \frac{\textrm{ln}(\gamma)}{\textrm{ln}(1-\epsilon)} \;.
\end{equation*} }
}

\vspace{1mm}

\noindent 2) Fig.~\ref{fig:distrib_Kopt_req1}: As a result of relaxing Req.~2 (from worst case to CVAR criterion), the Req.~1 is considerably improved: all metrics were reduced by~$\sim$30\% with respect to Section~\ref{sec:4_deter}.

\begin{table}[ht!]
    \centering
   \caption{Performance metrics for the deterministic (det.) and stochastic (sto.) approaches.}
    \label{table:results}
    \normalsize
   \begin{tabular}{| l || c | c | c | c | c |}
     \hline 
     Req.1 & Nominal & Mean & $\beta$-VAR & $\beta$-CVAR & Worst \\ \hline \hline
      det. & 1.02 & 1.03 & 1.09 & 1.10 & 1.17 \\
      sto. & 0.69 & 0.70 & 0.73 & 0.74 & 0.79 \\ \hline \hline \hline
     Req.2 &  & &  & &  \\ \hline 
      det. & 0.90 & 0.87 & 0.91 & 0.92 & 0.93 \\
      sto. & 0.87 & 0.79 & 0.87 & 0.98 & 2.05 \\ \hline \hline \hline
     Req.3 &  & &  &  & \\ \hline 
      det. & 1.00 & 1.00 & 1.00 & 1.00 & 1.00 \\
      sto. & 1.00 & 1.00 & 1.00 & 1.00 & 1.05 \\ \hline
   \end{tabular}
 \end{table}